\documentclass[a4paper, French , 12pt ,oneside  , leqno ] {article}

\usepackage[latin1]{inputenc}
\usepackage[TS1,T1]{fontenc}
\usepackage{ae}
\usepackage{aeguill}
\usepackage[francais]{babel}

\usepackage{amssymb}

\usepackage{amsmath}
\usepackage{amsthm}
\usepackage{amscd}

{}

\usepackage[TS1,T1]{fontenc}
\usepackage{ae}
\usepackage{aeguill}
\usepackage[francais]{babel}

\numberwithin{equation}{section}
 
\newtheorem{prop}{Proposition}[section]
\newtheorem{cor}[prop]{Corollaire}

\newtheorem{lem}[prop]{Lemme}
\newtheorem{thm}[prop]{Th{\'e}or{\`e}me}
\theoremstyle{definition}

\newenvironment{Rem}{\noindent {\bf Remarque :} }{}

\newenvironment{nots}{\noindent {\bf Notations :} }{}

\newenvironment{preuve}{\noindent {\bf Preuve :} }{}

\newenvironment{preuve thm}{\noindent {\bf Preuve du th{\'e}or{\`e}me :} }{}
\newenvironment{preuve prop}{\noindent {\bf Preuve de la  proposition:} }{}
\newenvironment{R{\'e}s}{\noindent {\bf R{\'e}sum{\'e} :} }{}

\newenvironment{R{\'e}f}{\noindent {\bf R{\'e}f{\'e}rences:} }{}
\newenvironment{ab}{\noindent {\bf Abstract:} }{}
\newcommand{\email}{E-mail:}

\newcommand{\ds}{\displaystyle}
\newcommand{\mud}{\mu_{d_1}}
\newcommand{\mudn}{\mu_{d_n}}

\newcommand{\Z}{\mathbb{Z}}
\newcommand{\N}{\mathbb{N}}         
\newcommand{\C}{\mathbb{C}}
\newcommand{\R}{\mathbb{R}}

\def\P{\mathbb P} 

\newcommand{\e}{\mathbb E}
\def\md {\par \medskip}
\newcommand{\di}{\diamondsuit}
\def\pn {\par \noindent}

\begin{document}  
\title {\bf Densit{\'e} d'{\'e}tat surfacique pour une classe d'op{\'e}rateurs de
  Schr{\"o}dinger du type {\`a} N-corps }
 
\author{Boutheina {\sc Souabni } }
\date{}
\maketitle

{\footnotesize{\ab We are interested in quantum systems composed of a finite number  of
 particles and described by Hamiltonians which are random
 Schr{\"o}dinger operators $H^{\omega
}:=-\Delta + V^{\omega} $ on
 $L^2(X)$, where $X$ is a finite dimensional Euclidean space and
 $\Delta$ is the  Laplace-Beltrami operator on $X$. We consider $X$
 as the configuration space of the system and we assume
 that $\{X_n\}_{1\leqslant n \leqslant N_0}$ is a family of linear
 subspaces of $X$. The orthogonal complement of $X_n$ in $X$ is denoted $
 X^{n}$ and is considered as the configuration space of a subsystem. We
 assume that $V^{\omega}$ is a sum of potentials  $v_n^{\omega}: X \longrightarrow \R,\quad 1\leqslant n \leqslant N_0,$ which are ergodic with respect the translation group of
 $X_n$ and which are rapidly decaying in any
direction of $X^{n}.$ The aim of this paper is to show the existence of
a thermodynamical limit. This limit defines an object which is a type of
a
the integrated density of states in the case of two body systems.\md\pn 


}}

\setcounter{section}{0}
\section{Introduction}
On s'int{\'e}resse {\`a} un syst{\`e}me quantique form{\'e} d'un nombre fini de
particules dont l'hamiltonien est d{\'e}fini  par un op{\'e}rateur de
Schr{\"o}dinger al{\'e}atoire $H^{\omega}:=-\Delta + V^{\omega} $ dans l'espace de Hilbert $L^2
(X)$, o{\`u} $X $ est un espace euclidien de dimension finie et $\Delta$
est l'op{\'e}rateur de Laplace-Beltrami. On consid{\`e}re
$X$
comme l'espace de configuration des particules et on fixe une famille
$\{ X_n\}_{0\leqslant n \leqslant N_0} $ de sous-espaces de $X$. Le
suppl{\'e}mentaire orthogonale de $ X_n$ dans $X$ est not{\'e} $ X^{n}$, 
il est consid{\'e}r{\'e}
comme l'espace de configuration d'un sous-syst{\`e}me du syst{\`e}me initial
des particules, pour plus de d{\'e}tails sur le probl{\`e}me {\`a} N-corps
voir \cite{anne}. \md \pn  On suppose que $V^{\omega}$ est une somme des
potentiels $v_n^{\omega}: X \longrightarrow \R $ ergodiques par rapport
au groupe des translations de $ X_n$ et d{\'e}croissantes rapidement en direction
de $X^n$.\md \pn Le but de cet article est de montrer l'existence d'une limite
thermodynamique. Cette limite d{\'e}finit un
objet du type de la densit{\'e} d'{\'e}tat qui correspond {\`a} la densit{\'e} d'{\'e}tat
surfacique dans le cas du syst{\`e}me {\`a} 2 corps .\md \pn      
Plus pr{\'e}cisement on suppose  $X = (\R^d , <.,.>)$  muni du produit
scalaire $$<x,y> :=\ds
\sum_{1\leqslant j,k\leqslant d } g_{jk}x_j y_k $$ et$$\Delta :=\ds \sum_{1\leqslant j,k\leqslant d }
g^{jk}\frac{\partial^2}{\partial x_j \partial x_k},$$o{\`u} $\
(g_{jk})$ est une  matrice
sym{\'e}trique strictement positive et  $(g^{jk})$ son inverse. Par
cons{\'e}quent :
\begin{equation}\label{a} 
H^{\omega} =
- \Delta + \sum_{n=0}^{N_0} v_n ^\omega (x) .
\end{equation}
On suppose que $( \Omega, P) $ est un  espace de probabilit{\'e}. Pour tout $z \in X $
on consid{\`e}re $\tau_z : \Omega \rightarrow  \Omega $ mesurable, pr{\'e}servant
la mesure  tel que  $v_n^\omega (x+z)= v_n^{\tau_z\omega}(x)$ pour tout $z \in
X_n$. On suppose que $(\tau_z)_{z\in X_n }$ est ergodique pour
$n=1,2,....,N_0$. On suppose pour tout $\gamma>0$ il existe une
constante $ C_{\gamma}>0$ telle que $$\mid v_n^{\omega}(x)\mid\leqslant
C_{\gamma} (1+\mid \pi^n x\mid)^{-\gamma},$$o{\`u} $\pi^n:X\longrightarrow
X$ d{\'e}signe la projection  sur $X^n$ le long de $X_n .$ 

\begin{nots}
On pose  $X_0 = X, \ H_0 = -\Delta + v_0(x),$\md \pn et $d_n = dim X_n$ avec $d=d_0\geqslant d_1\geqslant d_2
....\geqslant d_{N_0
}\geqslant 0
.$ \md \pn 
On d{\'e}signe par 
$A_L = B_X(0,L) $ la boule  dans $
X$ de centre 0 et de rayon L et $\chi_{A_L}$ d{\'e}signe sa fonction caract{\'e}ristique. 
\end{nots}

Pour tout $f \in C_0^\infty ({\R})$ on d{\'e}finit : 
\begin{equation}\label{b}  
\nu_s^L :=
\frac{1}{\mud L^{d_1}}\ tr \{ \chi_{A_L}(f(H)-f(H_0))\}.
\end{equation} o{\`u} $\mu_{d_k}$ le volume de la boule unit{\'e} habituelle  dans
$\R^{d_k} .$\md \pn
 La justification du fait que la quantit{\'e}  $\nu_s^L $ soit bien d{\'e}finie 
se trouve dans la section \ref{quatre} (voir proposition $(\ref{paaa}))$.\md
 Le\  r{\'e}sultat principal de cette partie est le
  th{\'e}or{\`e}me suivant :
\begin{thm}\label{th} {\bf Existence de la densit{\'e} d'{\'e}tat surfacique}\md \pn
On suppose que $ f \in C^{\infty}_0 ({\R}).$   Alors $\displaystyle 
\lim_{ L \rightarrow
  \infty}\  \nu_s^L(f)  = \nu_s(f)$ existe P-presque partout et elle
est non
  al{\'e}atoire .
\end{thm}
Si $ f \in C^{\infty}_0({\R})$ v{\'e}rifie  $f\geqslant 0$ et
  supp$f\subset{\R}\backslash \sigma (H_0)$ 
alors,$$\nu_{s}^{L}(f) = \frac{1}{\mud L^{d_1}}tr\{\chi_{A_{L}}f(H)\}\geqslant 0.$$
Ainsi la fonctionnelle $\nu_{s}^{L}$\ est  positive,  et $\nu_s$\
l'est aussi. D'apr{\`e}s le th{\'e}or{\`e}me de repr{\'e}sentation de
Riesz on obtient,
\begin{cor}
La restriction de $ \nu_s \  \grave{a} \ {\R} \backslash \sigma (H_0 )
$ est une mesure positive.
\end{cor}
\md
Dans la th{\'e}orie spectrale de l'op{\'e}rateur de Schr{\"o}dinger (continu ou discret) il y a deux
classes d'op{\'e}rateurs qui ont {\'e}t{\'e} tr{\`e}s {\'e}tudi{\'e}es. La premi{\`e}re classe est
celle des op{\'e}rateurs dont le potentiel est d{\'e}croissant {\`a} l'infini,
elle fait l'objet de la th{\'e}orie de diffusion. L'une des notions importantes de
cette th{\'e}orie est la fonction de d{\'e}placement spectral,
introduite par I. Lifshitz et M. Krein $\cite{Birman}$. La seconde classe est
form{\'e}e par des op{\'e}rateurs dont le potentiel est par
exemple p{\'e}riodique, presque p{\'e}riodique, quasi p{\'e}riodique ou al{\'e}atoire
ergodique (voir par exemple $\cite{Avron},\cite{Carmona},\cite{Nakumura},\cite{Pastur},\cite{Simon})$. Dans ce cas il
y a une caract{\'e}ristique importante {\`a} savoir la densit{\'e} d'{\'e}tat int{\'e}gr{\'e}e qui d{\'e}tecte le spectre de H.\\
Il y a  une autre classe d'op{\'e}rateurs, qui est le cas interm{\'e}diaire, o{\`u} le
champ al{\'e}atoire ergodique est port{\'e} par une surface de l'espace tout
entier,  consid{\'e}r{\'e} par Chahrour $\cite{Chahrour}$ pour le
cas discret, et par Schr{\"o}der, Simon, Kirsch, English dans le cas
continu et discret $(\cite{Englishk},\cite{English}).$ Ces travaux ont {\'e}t{\'e} consacr{\'e}s
{\`a} la construction de l'objet appel{\'e} la densit{\'e} d'{\'e}tat surfacique qui d{\'e}tecte le spectre de $H$.\\
C'est pourquoi nous nous concentrons ici sur le cas de Schr{\"o}dinger
continu d{\'e}fini ci dessus, nous prouvons (Th{\'e}or{\`e}me $\ref{th}$) qu'une
quantit{\'e} analogue adopt{\'e}e {\`a} notre situation et 
que nous appelons aussi `` la densit{\'e} d'{\'e}tat surfacique'' existe en tant que distribution.\\
Les d{\'e}monstrations  utilisent les commutateurs convenablement exploit{\'e}s, la propri{\'e}t{\'e} de la norme trace et des propri{\'e}t{\'e}s de l'op{\'e}rateur
$e^{-tH}$ .\md
\md
\begin{Rem}
On note
 $$ H_n = H_0+ v_n,\quad \nu_{s,L}^n =
\frac{1}{\mudn L^{d_n}}\ tr \{ \chi_{A_{L}}(f(H_n)-f(H_0))\} .$$ 
Alors $\nu_s^n(f)=\ds \lim_{ L \rightarrow
  \infty}\  \nu_{s,L}^n(f)$ existe et le raisonnement
pr{\'e}sent{\'e}  dans cet article permet de lier  le
supp$\nu_s$ et supp$\nu_s^n$:$$ {\rm supp} (\nu_s) \backslash
\sigma(H_0)= \mathop \cup_{n=1}^{N_1} {\rm supp} (\nu_s^n)\backslash \sigma(H_0)
,$$avec $N_1= \#{\cal K}$ et ${\cal K}=\{n \in
\N, \ d_n=d_1\}$. (On peut dire que pour $ 1\leqslant n\leqslant N_1$ le
potentiel $ v_n $ est  une interaction {\`a} 2 corps). Or on a toujours    $${\rm supp} (\nu_s) \backslash\sigma(H_0)
\subset\sigma(H)\backslash\sigma(H_0),$$   par cons{\'e}quent on
trouve $$\mathop \cup_{n\in {\cal K}} {\rm supp} (\nu_s^n)\backslash \sigma(H_0)\subset\sigma(H)\backslash\sigma(H_0)
.$$ Si on suppose de
plus que pour tout $ n \in \{1,...,N_0\}$, la fonction $ v_n$
est continue et presque
p{\'e}riodique par rapport {\`a} $\pi_n x$,  on a alors un r{\'e}sultat analogue {\`a} celui de Chahrour
$\cite{Chahrour}$ sur le spectre :$${\rm supp} (\nu_s^n) \backslash
\sigma(H_0)=  \sigma(H_n)\backslash \sigma(H_0)$$
La preuve de ce r{\'e}sultat sera donn{\'e}e dans
un travail ult{\'e}rieur.
\end{Rem}
\setcounter{section}{1}
\section{Preuve du Th{\'e}or{\`e}me \ref{th}}\label{un}
La d{\'e}monstration de ce th{\'e}or{\`e}me s'appuie sur les Propositions
$\ref{pa}, \ref{pe}$ et $\ref{pd}$. Pour les {\'e}noncer nous avons besoin des 
notations suivantes: \md \pn Posons $\pi_n:
X\longrightarrow X $ la projection   sur $X_n$  avec Im$\pi_n
= X_n$,  Ker$\pi_n = X^n$ et  $\pi^n =
I_{X}-\pi_n . $\md\pn 
Pour $ p \geqslant 1$ on note $  \mid \mid A
\mid \mid_{B_p}=(tr \mid A \mid^p)^{\frac{1}{p}}$ et
on dit que A est
un op{\'e}rateur {\`a} trace si $$ \mid\mid A \mid\mid_{B_1}<\infty.$$
 Pour $k \in \Z^d,$ on note $ \chi_k $  la fonction caract{\'e}ristique du
cube  $ {\cal C}_k = k + [-\frac{1}{2} ,\frac{1}{2}]^d $ et \md\pn $ \mid k \mid_1 = k_1+.....+k_d. $ 
\begin{prop} \label{pa}Soit $\alpha\in ]0,\frac{1}{d²}[$, pour tout $n \in\{1,.....,N_0\}$ on note :
$$A_{L,n} := \{x\in A_L / \mid \pi^n x\mid\leqslant L^{\alpha} \}.$$
Alors pour  toute $ f \in C_0^{\infty}(\R^d), $ on a: 
\begin{eqnarray}\label{c}\displaystyle 
\lim_{ L \rightarrow
  \infty}\ \frac{1}{\mud L^{d_1}} tr \{ \chi_{A_{L,n}}(f(H)-f(H_n)) \} =
0  \ \  \  .\end{eqnarray} 
\end{prop}
Pour montrer cette proposition nous utilisons les lemmes suivants:
\begin{lem}\label{pb}
  Pour tout $N \in {\N}$ il existe
 une constante $c'_{N}>0$ telle que: $$\mid \mid \chi_m e^{-tH} \chi_k
 \mid \mid _{B_1}\leqslant \frac{c'_{N}}{t^{d}(1+\mid k-m \mid^N)},$$
 si $0<t\leqslant1$. Plus
 g{\'e}n{\'e}ralement pour tout  $\chi \in C_0^{\infty}({{\R}^d}),$  on a   
$$ \mid \mid \chi e^{-tH} \mid \mid_{B_{1}} \leqslant C t^{-d}\mu(\chi),$$ o{\`u}
$\mu(\chi) =\#\{ j\in\Z^d:\  {\rm supp}\chi\cap {\cal C}_j\neq\emptyset \}
$ et ${\cal C}_j=j+[ \frac{-1}{2}, \frac{1}{2}]^d.$
\end{lem}
\begin{lem}\label{pc'}
Pour tout $\epsilon>0$ et $ 0<t\leqslant 1$ on a 
$$\lim_{ L \rightarrow
  \infty}\ \frac{1}{\mud L^{d_1}} \mid\mid  (1-\chi_{\epsilon})\chi_{A_{L,n}}(e^{-tH}-e^{-tH_n}) \mid\mid_{B_1}
=0,$$  o{\`u}  $\chi_{\epsilon}$ est la fonction caract{\'e}ristique de $$\{x\in
A_{L,n}/ \ds\min_{1\leqslant n\leqslant N_0}\mid\pi^n x\mid\leqslant L^{\epsilon}\}=A_{L,n}\backslash \cap_{n=1}^{N_0}\{\mid\pi^n x\mid>L^{\epsilon}\}.$$ 
\end{lem}\pn Pour la preuve des deux lemmes voir paragraphe \ref{quatre}.
\begin{lem}\label{pc}
Pour tout $n\in\{1,....,N_0\},$ on a:
\begin{equation}\label {c1}  
\lim_{ L \rightarrow
  \infty}\ \frac{1}{\mud L^{d_1}} tr \{ \chi_{A_{L,n}}(e^{-H_n}-e^{-H}) \} =
0  \ \  \  .
\end{equation}
\end{lem}
\begin{preuve}  
On va utiliser la commutation avec des fonctions-troncatures. Soient
$\alpha\in ]0,\frac{1}{d²}[,$ on d{\'e}finit $w_{L^{\alpha}} $  par
:$$w_{L^{\alpha}} :=
\chi_{A_{3L,n} }*\gamma_{L^{\alpha}}(x),  $$ o{\`u} $$ \gamma_L(x) =
L^{-d} \gamma_1(\frac{x}{L}) \ ,\gamma_1 \in C_0^{\infty}(X)\ {\rm
et}
\ \int \gamma_1(x) dx =1 .$$ 
Il est clair que 
 $$\mid\partial^{\beta}_{ x} w_{L^{\alpha}}(x)\mid \leqslant c_{\beta}L^{-\mid\beta\mid
   \alpha},$$ pour tout $\beta \in \N^{d}$  et  choisissant $\gamma_1$
   convenablement on peut assurer $$w_{L^{\alpha}} (x)=
 0 \quad{\rm si}\quad  \mid \pi^n x \mid \geqslant
 4L^{\alpha}\quad{\rm et}\quad w_{L^{\alpha}}(x)=1 \quad
  {\rm si}\quad\mid \pi^n x \mid \leqslant 2L^{\alpha}. $$
De la m{\^e}me fa{\c c}on on d{\'e}finit $\widetilde{w}_L$ qui v{\'e}rifie  $$\mid\partial^{\beta}_{ x}\widetilde {w_{L}}(x)\mid \leqslant c_{\beta}L^{-\mid\beta\mid
   },$$ pour tout $\beta \in \N^{d}$  et
  $$ \widetilde{w}_L(x)=0 \quad {\rm si} \quad\mid \pi_n x \mid \geqslant
  4L \quad{\rm et}\quad
  \widetilde{w}_L(x)= 1\quad {\rm si} \quad\mid \pi_n x \mid \leqslant
  2L .$$ 
  Dans la suite on pose $w_L (x)= \widetilde{w}_{L}(x) w_{L^{\alpha}} (x).$
En vertu du Lemme \ref{pc'},  montrer  $(\ref{c1})$ {\'e}quivaut {\`a} montrer :$$\lim_{ L \rightarrow
  \infty}\ \frac{1}{\mud L^{d_1}} \mid\mid \chi_{\epsilon}\chi_{A_{L,n}}(e^{-H_n}-e^{-H}) \mid\mid_{B_1} =
0,$$ et compte tenu de  $\chi_{A_{L,n}}w_L(x)= \chi_{A_{L,n}}(x),$ on
a$$ \mid\mid
\chi_{\epsilon}\chi_{A_{L,n}}(e^{-H_n}-e^{-H}) \mid\mid_{B_1}\leqslant
\mid\mid\chi_{\epsilon}( w_L e^{-H_n}-
e^{-H}w_L)\mid\mid_{B_1}+\mid\mid\chi_{A_{L,n}}e^{-H}(1-w_L)\mid\mid_{B_1}.$$En vertue du Lemme \ref{pddd}, et vue que
dist(supp$\chi_{A_{L,n}}$,supp$(1-w_L))\geqslant cL^{\alpha},$ 
pour tout $N>0$ il existe une constante $C_N>0$ telle que
$$\mid\mid\chi_{A_{L,n}}e^{-H}(1-w_L)\mid\mid_{B_1}\leqslant c_N
L^{-N}.  $$Ainsi pour montrer (\ref{c1}) il suffit de montrer que: 
\begin{equation}\label {dd}
\lim_{ L \rightarrow
  \infty}\ \frac{1}{\mud L^{d_1}} \mid \mid \chi_{\epsilon}(w_{L}e^{-H_n}-e^{-H}w_L) 
\mid \mid_{B_1}=0. 
\end{equation}
Pour simplifier la notation  on prend n = 1. On a alors
\begin{eqnarray}  w_{L}e^{-H_1}-e^{-H}w_L  &=& \int^t_0 \frac{d}{dr}e^{-(t-r)H}w_L e^{-rH_1}dr \nonumber\\  & =&\int^t_0 e^{-(t-r)H}[H ,w_L]e^{-rH_1}+ e^{-(t-r)H}w_L\sum_{k=2}^{N_0} v_k e^{-rH_1}dr.\nonumber
\end{eqnarray} 
\pn  Pour  montrer $$\lim_{ L\rightarrow\infty}\ \frac{1}{\mud L^{d_1}}
\mid\mid\int_0^t \chi_{\epsilon}e^{-(t-r)H}w_L\sum_{k=2}^{N_0} v_k  e^{-rH_1} dr \mid\mid_{B_1}= 0,$$
on pose $$J_1 =\frac{1}{\mud L^{d_1}}\int _0^{\frac{t}{2}}\chi_{\epsilon} e^{-(t-r)H}
w_L\sum_{k=2}^{N_0} v_k  e^{-rH_1} dr, $$ et $$J_2= \frac{1}{\mud L^{d_1}}\int _{\frac{t}{2}}^t \chi_{\epsilon}e^{-(t-r)H}
w_L\sum_{k=2}^{N_0} v_k e^{-rH_1} dr.$$ 
Le Corollaire $\ref{pfff}$ (voir Appendice) assure l'existence d' une constante $c>0$ telle que  l'on ait :
\begin{equation}\label{c'}
\mid\mid\sum_{k=2}^{N_0} v_k \ w_L\  e^{-rH}  \mid
\mid_{B_1}\leqslant \frac{c L^{d_1-1+\alpha d}}{r^{d}} 
\end{equation} et
\begin{equation}\label{b'}
\mid\mid e^{-(t-r)H}w_L\sum_{k=2}^{N_0} v_k \mid\mid_{B_1}
\leqslant\frac{cL^{d_1-1+\alpha d}}{(t-r)^{d} }.
\end{equation}
On obtient 
\begin{eqnarray}
\mid \mid J_1\mid\mid_{B_1} &\leqslant&\frac{1}{\mud L^{d_1}}\int
_0^{\frac{t}{2}} \mid\mid
\chi_{\epsilon}e^{-(t-r)H}w_L\sum_{k=2}^{N_0} v_k
\mid\mid_{B_1}\mid\mid e^{-rH_1}\mid\mid dr \nonumber\\
&\leqslant&\int_0^{\frac{t}{2}} \frac{c}{L^{1-\alpha d}(t-r)^{d}} dr \to 0\quad
{\rm si}\quad  L\  \to \infty.\nonumber 
\end{eqnarray}
Il en est de m{\^e}me pour $J_2,$ d'o{\`u}  $$\mid\mid
J_2\mid\mid_{B_1} \leqslant \frac{1}{L^{1-\alpha d}}\int _{\frac{t}{2}}^t
\frac{c}{r^{d}}dr \to 0 \quad{\rm si}\quad  L\quad  \to \infty .$$
Il reste {\`a} montrer  $$\lim_{ L \rightarrow
  \infty}\mid\mid \frac{1}{\mud L^{d_1}}\int^t_0
 \chi_{\epsilon}e^{-(t-r)H}[H,w_L]e^{-rH_1}dr\mid\mid_{B_1} = 0. $$
Soit $$I_1 =\int^{\frac{t}{2}}_0 \chi_{\epsilon}e^{-(t-r)H}[H,w_L]e^{-rH_1}\quad dr.$$
Pour calculer  $[H,w_L],$ on introduit 
  $$p(x,\xi)= \ds \sum _{1\leqslant (j,k)
\leqslant d} g^{jk}\xi_j \xi_k+  \sum_{n=0}^{N_0}v_n^{\omega}(x),$$ 
 on remarque que  la composition $\widetilde{p}(x,D) :=  p(x,D)
 w_L(x) $, (voir
\cite{Robert}), poss{\`e}de le symbole  
\begin{eqnarray}
\widetilde{p}(x,\xi) & := &  \sum_{\alpha} \frac{1}{i^{\mid \alpha \mid} \alpha!}[(\partial^{\alpha}_{\xi}p)(x,\xi)\
\partial^{\alpha}_xw_L(x)]\nonumber.
\end{eqnarray}
On obtient :
$$[H,w_L]=\sum_{1\leqslant (j,k) \leqslant d} i g^{jk}\frac{1}{L^{\alpha}}
W_{L,j}(D_{k}+D_{j}) +\sum_{1\leqslant k\leqslant d}g^{kk}\frac{\widetilde{W}_{L,k}}{L^{2\alpha}},$$ o{\`u} $$W_{L,j}(x) = L^{\alpha}\partial_{j} w_L(x),\quad
\widetilde{W_{L,j}}(x) = L^{2\alpha}\partial_{j^2}^2
w_L(x)$$ sont uniform{\'e}ment born{\'e}es par rapport {\`a} L.
Par cons{\'e}quent, 
\begin{eqnarray}\label{b''}
\mid\mid
I_1\mid\mid_{B_1}\!\!\!\!&\leqslant&\!\!\!\!\sum_{j=1}^{d}\frac{c}{\mud
  L^{d_1+\alpha}}\!\!\!\int^{\frac{t}{2}}_0\!\!\!\mid\mid
\!\!\chi_{\epsilon}e^{-(t-r)H}W_{L,j}\!\!\mid\mid_{B_1}\mid\mid\! D_j(H_1+c)^{-1/2}\mid\mid\ \mid\mid\!(H_1+c)^{1/2}e^{-rH_1}\mid\mid dr\nonumber\\
& &+\sum_{j=1}^{d}\frac{c}{\mud
  L^{d_1+\alpha}}\int^{\frac{t}{2}}_0\mid\mid
\chi_{\epsilon}e^{-(t-r)H}\widetilde{W}_{L,j}\mid\mid_{B_1}\mid\mid e^{-rH_1}\mid\mid
dr.
\end{eqnarray}
Comme pour tout $\epsilon>0$ il existe $c>0$ telle que
$$\mu(\chi_{\epsilon})\leqslant cL^{\epsilon(d-d_1)+d_1}, $$ on peut
conclure par le Lemme \ref{pb} que $$\mid\mid \chi_{\epsilon}e^{-(t-r)H}
\widetilde{W}_{L,j}\mid\mid_{B_1}\leqslant c L^{\epsilon (d-d_1)+
  d_1}(t-r)^{-d}$$ et  $$\mid\mid \chi_{\epsilon}e^{-(t-r)H}
W_{L,j}\mid\mid_{B_1}\leqslant c L^{\epsilon
  (d-d_1)+d_1}(t-r)^{-d}.$$
Ainsi l'expresion (\ref{b''}) est major{\'e}e par 
$$\int^{\frac{t}{2}}_0\frac{C}{\sqrt{r}}\frac{1}{L^{\alpha-\epsilon
  (d-d_1)}}dr,$$et en choisissant  $\epsilon>0$ suffisament
petit on obtient: $$\lim_{L\to\infty}\mid\mid I_1\mid\mid_{B_1}=0.$$
De m{\^e}me, soit $$I_2 =\frac{1}{\mud L^{d_1}}\int_{\frac{t}{2}}^t
\chi_{\epsilon}e^{-(t-r)H}[H,w_L ]e^{-rH_1}dr,$$ et pareillement 
\begin{eqnarray}
\mid\mid I_2\mid\mid_{B_1}\!\!\!\! &\leqslant&\!\!\!\!\sum_{j=1}^{d}\!\! \frac{c}{\mud L^{d_1+\alpha}}\!\!\int_{\frac{t}{2}}
^t\!\!\mid\mid \chi_{\epsilon}W_{L,j}e^{-rH_1}\mid\mid_{B_1}\mid\mid D_j (H +c)^{-1/2}\mid\mid \ \mid \mid
(H +c)^{1/2}e^{-(t-r)H}\mid\mid dr\nonumber\\ & &+
\sum_{j=1}^{d}\frac{c}{\mud L^{d_1+\alpha}}\int_{\frac{t}{2}}
^t\mid\mid\chi_{\epsilon}\widetilde{W}_{L,j} e^{-rH_1}\mid\mid_{B_1} \ \mid\mid e^{-(t-r)H}\mid\mid dr\nonumber\\ &\leqslant& \int_{\frac{t}{2}}^t
\frac{C}{\sqrt{t-r}}\frac{1}{L^{\alpha-\epsilon
  (d-d_1)}} dr \to 0 \quad {\rm si}\quad  L\  \to \infty. \nonumber\\
\nonumber
\end{eqnarray}
Ainsi  pour $ L\
\to \infty $ on  a : $$\mid \mid I\mid \mid_{B_1} = \frac{1}{\mud L^{d_1}} \mid \mid \int^t _0
\chi_{\epsilon}e^{-(t-r)H}[H,w_L]e^{-rH_1}dr\mid \mid_{B_1} \leqslant\  \mid\mid
I_1\mid\mid_{B_1} +\mid\mid I_2 \mid\mid_{B_1} \to 0 . \di $$
\end{preuve}
\pn{\bf Preuve de la Proposition \ref{pa}:} On suppose
$\alpha<\frac{1}{d²}$ alors on peut trouver $\alpha'$ verifiant
$\alpha d <\alpha'<\frac{1}{d}.$ Pour simplifier la
d{\'e}monstration on prend $n =1.$ Alors pour toute $f\in
C^{\infty}_0 (\R) $  on peut {\'e}crire:$$f(x) = e^{-x} g(x),$$ o{\`u}
  $g \in C_0^{\infty}(\R).$ Alors: 
\begin{eqnarray}
\frac{1}{\mud L^{d_1}} tr\{\chi_{A_{L,1}}(f(H)-f(H_1))
\} & = &\frac{1}{\mud L^{d_1}}
tr\{\chi_{A_{L,1}}(e^{-H}g(H)-e^{-H_1}g(H_1))\}\nonumber\\ &=&\widetilde{I_1}+\widetilde{I_2},\nonumber 
\end{eqnarray}
o{\`u}  $$\widetilde{I_1}= \frac{1}{\mud L^{d_1}}
tr\{\chi_{A_{L,1}}(e^{-H}-e^{-H_1})g(H)\}\ ,$$ et   $$\widetilde{I_2}=\frac{1}{\mud L^{d_1}}
tr\{\chi_{A_{L,1}}e^{-H_1}(g(H)-g(H_1))\}.$$ 
En vertu du Lemme \ref{pc'} si on prend $n=1$, pour
montrer $\ds\lim_{L\to \infty}\widetilde{I_1}=0$ il suffit de prouver
$\ds\lim_{L\to\infty}\widetilde{\widetilde{I_1}}=0 $ o{\`u}$$\widetilde{\widetilde{I_1}}=\frac{1}{\mud L^{d_1}}
\mid\mid\chi_{\epsilon}\chi_{A_{L,1}}(e^{-H}-e^{-H_1})g(H)\mid\mid_{B_1}.$$
De la m{\^e}me fa{\c c}on que dans la preuve du Lemme \ref{pc} en utilisant le
fait que dist(supp$\chi_{A_{L,1}}$, supp$(1-w_L))\geqslant
cL^{\alpha}$ pour tout $N>0$ il existe une constant $C_N>$ telle que     :
\begin{equation}\label{cc}
 \mid\widetilde {\widetilde {I_1}}\mid \leqslant  \frac{1}{\mud L^{d_1}}( \mid \mid
\chi_{\epsilon}(w_{L}e^{-H}-e^{-H_1}w_L)\mid \mid_{B_1}+c_N L^{-N}) \mid \mid g(H)
\mid \mid,
\end{equation}
Or d'apr{\`e}s l'expression (\ref{dd}) le premier terme de (\ref{cc}) tend vers z{\'e}ro
et vue que  $g(H)$ est born{\'e} on a donc
$$\lim_{L\to\infty} \mid\widetilde {\widetilde {I_1}}\mid =0.$$ 
Montrons maintenant que $$   \lim_{L \rightarrow \infty} \mid \widetilde{I_2} \mid =0.$$
Soit $$A_{L,1}'=  \{ x \in A_L /
 \mid  \pi^1 x \mid \leqslant L^{\alpha' } \}.$$ 
En utilisant la convolution avec $\gamma_1$ de mani{\`e}re semblable {\`a}
celle 
consid{\'e}r{\'e}e  au d{\'e}but de la
preuve du Lemme \ref{pc}, on peut d{\'e}finir    $w_{L,1}' $ v{\'e}rifiant  : $$
\mid\partial_x^{\beta}(w_{L,1}')(x)\mid\leqslant C_{\beta}
L^{-\alpha'\mid\beta\mid},$$  pour  tout $\beta
\in \N^d $ et  
$${\rm dist}(A_{L,1},{\rm supp}(1- w_{L,1}')) \geqslant cL^{\alpha}.$$ 
Ainsi pour montrer que $\ds\lim_{L\to\infty}\widetilde {I_2}=0$ il suffit
de montrer que $\ds\lim_{L\to\infty}\widetilde{\widetilde {I_2}}=0$ o{\`u} 
$$\widetilde{\widetilde {I_2}}=
\frac{1}{\mud L^{d_1}}\mid\mid \chi_{A_{L,1}}e^{-H_1}( w_{L,1}'
g(H)-g(H_1)w_{L,1}')\mid \mid_{B_1}.$$En effet, en suivant le m{\^e}me
proc{\'e}dure que  $\widetilde{\widetilde {I_1}},$ pour tout $N>0$ il
existe une constante $C_N>0$ telle que 
\begin{eqnarray}
\mid \widetilde{I_2} \mid &= & \mid \frac{1}{\mud L^{d_1}}tr\{
\chi_{A_{L,1}}w_{L,1}' e^{-H_1}(g(H)-g(H_1))\}\mid\nonumber\\
&\leqslant&\frac{1}{\mud L^{d_1}}\widetilde{\widetilde {I_2}} +C_NL^{-N}.\nonumber
\end{eqnarray}
En utilisant la formule de Helffer-Sj{\"o}strand  :\md \pn 
si $f \in C_0^{\infty}({\R}),$ il existe $f_1\in
C_0^{\infty}(\C)$ telle que $$\mid \partial_{\bar
    z}(f_1) \mid \leqslant c_N\mid Imz \mid^N,$$  pour tout
  $N\in \N $ et $$f(H) = c\int_{K^2} \partial
  _{\bar z} f_1 (a+ib)(H-(a+ib))^{-1}da db,$$ o{\`u} $ {\rm
    supp}(f_1)\subset K^2.$ Si on pose $R_z(H)
=(H-z)^{-1},$ on obtient  
\begin{eqnarray}
\mid\widetilde{ \widetilde{I_2} }\mid &\leqslant&  \frac{c}{\mud L^{d_1}}\mid\mid \chi_{A_{L,1}}e^{-H_1}\int_{K^2}\partial_{\bar
    z}f_1(a+ib)(w_{L,1}' R_z(H)-R_z(H_1)w_{L,1}')dadb\mid
\mid_{B_1}\nonumber\\ &\leqslant&\frac{C}{\mud L^{d_1}}\int_{K^2} 
  \sup_{z \in \C \setminus \R}\{\mid\mid\chi_{A_{L,1}}e^{-H_1}( w_{L,1}' R_z(H)-R_z(H_1)w_{L,1}')\mid\mid_{B_1}\ \mid\partial_{\bar z}f_1\mid \})dadb.\nonumber
\end{eqnarray}
Or 
\begin{eqnarray}
\widetilde I_{(2,z)}=w_{L,1}' R_z(H)-R_z(H_1)w_{L,1}'\!\!\!&\!\!\!=\!\!\!&R_z(H_1)((H_1-z)w_{L,1}'-w_{L,1}'(H-z))R_z(H)\nonumber\\
\!\!\!&=&\!\!\!R_z(H_1)([H_1,w_{L,1}']-\sum_{k=2}^{N_0}
w_{L,1}' v_k)R_z(H).\nonumber
\end{eqnarray}
Ainsi
\begin{eqnarray}\label{d'}
\mid\mid\!\!\chi_{A_{L,1}}e^{-H_1}\widetilde
I_{(2,z)}\mid\mid_{B_1}\!\!\!\!&\leqslant&\!\!\!\!\mid\mid\!\!\chi_{A_{L,1}}e^{-H_1}\!\!\mid\mid_{B_1}\mid\mid
R_z(H_1)[H_1,w_{L,1}']R_z(H)\mid\mid \nonumber\\\!\!\!\! & &\!\!\!\!+\mid\mid
R_z(H_1) \mid\mid \quad \mid\mid\chi_{A_{L,1}}e^{-H_1}\sum_{k=2}^{N_0}
w_{L,1}' v_k\mid\mid_{B_1}\mid\mid R_z(H)\mid\mid. 
\end{eqnarray}
Gr{\^a}ce {\`a} l'estimation du commutateur, en rempla{\c c}ant 
$L^{\alpha'}$ au lieu $L^{\alpha}$ dans la preuve du
Lemme 
\ref{pc}  et en vertu du Lemme \ref{pb} et en utilisant  $\mu(A_{L,1})\leqslant c
L^{d_1+\alpha d }$, le premier terme de  l'expression  (\ref{d'}) est major{\'e} par 
$$ \frac{c}{\mid Imz\mid^2}L^{d_1+\alpha d-\alpha'}.$$
D'apr{\`e}s l'expression (\ref{b'}) le deuxi{\`e}me terme est major{\'e} par 
$$\frac{c}{\mid Imz\mid^2}L^{d_1-1+\alpha'd}.$$
Ainsi en utilisant le fait que  $\alpha d<\alpha'<\frac{1}{d}$  on conclue:
$$\mid\widetilde { \widetilde I_2} \mid \leqslant c\ C L^{\alpha d-\alpha'
  }+c\ C L^{d_1-1+\alpha' d-d_1}\to 0 \quad {\rm si}\quad L\ 
\to \infty,$$ ce qui ach{\`e}ve la d{\'e}monstration.
 $\qquad \di$ \par \bigskip
En reprenant encore une fois la m{\^e}me convolution de $\gamma_1$, on peut d{\'e}finir une fonction lisse $ w_{L , \alpha}$
  {\`a} support compact telle que $$ w_{L , \alpha} \chi_{\{x\in A_L  \setminus\mid\pi^n x \mid\geqslant C
  L^{\alpha} \} }=\chi_{\{x\in A_L  \setminus\mid\pi^n x \mid\geqslant C
  L^{\alpha} \} }\quad {\rm et}\quad \mid\partial^{\beta} w_{L , \alpha}\mid \leqslant
c_{\beta}L^{ -\mid\beta\mid\alpha}$$pour tout $ \beta \in \N^d.$  Alors le
Lemme \ref{peee} assure que pour tout $N\in\N$ il existe une constante
$c_N>0$ telle que :
$$\mid\mid \sum_{n=1}^{N_0} v_n w_{L , \alpha} 
  e^{-tH}\mid\mid_{B_1} \leqslant \frac{c_N}{t^d L^{N}} $$   et
$$\mid\mid e^{-tH} w_{L , \alpha}\sum_{n=1}^{N_0} v_n
  \mid\mid_{B_1} \leqslant \frac{c_N}{t^d L^{N}}, $$pour tout
  $0<t\leqslant 1$.  Par cons{\'e}quent  suivant les m{\^e}mes {\'e}tapes que la d{\'e}monstration de la
Proposition \ref{pa}, on
obtient:
\begin{prop}\label{pa1}Pour toute $ f \in C_0^{\infty}(\R^d).$ Si  $J_L\subset\cap_{n=1}^{N_0}\{x\in A_L \setminus\mid\pi^n x \mid\geqslant C
  L^{\alpha} \}, $ alors :  
$$\displaystyle 
\lim_{ L \rightarrow
  \infty}\ \frac{1}{\mud L^{d_1}}tr\{ \chi_{J_L}(f(H)-f(H_0)) \} =
0  \ \  \ .$$  
\end{prop}
\begin{prop}\label{pe}
Si $f \in C_0^\infty ({\R})$ et $\nu_s^L(f)$ est  donn{\'e}e par $(\ref{b}),$
  alors
$$\displaystyle 
\lim_{ L \rightarrow \infty}\  \mid{  \nu_s^L(f) - \frac{1}{\mud L^{d_1}}
\sum_{n=1}^{N_0}tr\{ \chi_{ A_{L,n}}} (f(H_n) - f(H_0))\}\mid = 0 .$$ 
\end{prop}
\begin{preuve}En utilisant 
  :$$\chi_{\cup_{n=1}^{N_0}{A_{L,n}}}=\mathop\sum_{n=1}^{N_0}(\chi_{A_{L,n}} - \chi_{A_{L,n}\cap B_{L,n-1}})$$o{\`u}$$ B_{L,n}= \mathop\cup_{k=1}^{n} A_{L,k} \quad {\rm pour}\quad k\geqslant 1 \quad {\rm et} \quad B_{L,0} = \emptyset,$$
on peut  {\'e}crire
\begin{eqnarray}
 \chi_{A_L} &=&\chi_{
  A_{L}\backslash B_{L,N_0}}+\chi_{B_{L,N_0} } \nonumber \\ 
&=& \chi_{A_{L}\backslash B_{L,N_0}}+ \sum_{n=1}^{N_0}(\chi_{A_{L,n}}- \chi_{A_{L,n}\cap B_{L,n-1}}) . \nonumber
\end{eqnarray}
Il est clair que:
\begin{eqnarray} 
\nu_s^L &=& \frac{1}{\mud L^{d_1}}\
tr\{\chi_{A_L}(f(H)-f(H_0))\}\nonumber\\ &=&\frac{1}{\mud
  L^{d_1}}(\sum_{n=1}^{N_0}tr\{\chi_{ A_{L,n}}
(f(H)-f(H_0))\}-\sum_{n=1}^{N_0}tr\{\chi_{A_{L,n}\cap
  B_{L,n-1}}(f(H)-f(H_0)))\ \nonumber\\ & &+\frac{1}{\mud L^{d_1}}\ tr\{\chi_{A_L\backslash B_{L,N_0}}(f(H)-f(H_0))\}.\nonumber
\end{eqnarray}
On remarque que la  mesure de Lebesgue 
$$
 {\rm mes}\{A_{L,n}\cap B_{L,n}\} \leqslant c L^{d_1-1}, $$
 par cons{\'e}quent 
\begin{equation}\label{e}
 \mu(\chi_{A_{L,n}\cap B_{L,n}}) \leqslant c L^{d_1-1}.
\end{equation}
De la m{\^e}me fa{\c c}on que le Lemme \ref{pb} et la  Proposition \ref{paaa},
on peut conclure par 
  l'expression $(\ref{e})$ :
\begin{eqnarray}
\mid \frac{1}{\mud L^{d_1}}\sum_{n=1}^{N_0}tr\{\chi_{A_{L,n}\cap B_{L,n}}(f(H)-f(H_0))\}\mid 
&\leqslant&\frac{1}{\mud L^{d_1}} \sum_{n=1}^{N_0}c(f) cL^{d_1-1}\nonumber\\
&\leqslant & N_0 c(f)cL^{-1} \to 0 \quad {\rm si}\quad L\
\to \infty.\nonumber\\
\nonumber
\end{eqnarray}
En vertu de la Proposition $\ref{pa}$ on a, $$\lim_{ L \rightarrow
  \infty}\ \frac{1}{\mud L^{d_n}} tr \{ \chi_{A_{L,n}}
f(H_n)\} =\lim_{ L \rightarrow
  \infty}\ \frac{1}{\mud L^{d_n}} tr \{ \chi_{A_{L,n}}f(H) \} .$$
Ce qui donne $$\lim_{ L \rightarrow
  \infty}\sum_{n=1}^{N_0}\frac{1}{\mud L^{d_1}}\ tr \{
\chi_{A_{L,n}}(f(H_n)-f(H_0))\}=
\lim_{ L \rightarrow
  \infty} \sum_{n=1}^{N_0} \frac{1}{\mud L^{d_1}}\ tr \{
\chi_{A_{L,n}}(f(H)-f(H_0))\} .$$ 
Ainsi pour prouver la Proposition $\ref{pe}$, il suffit de montrer : 
 $$\displaystyle 
\lim_{ L \rightarrow \infty}\mid \frac{1}{\mud L^{d_1}}\ tr \{
\chi_{A_L\backslash A_{L,n}}(f(H)-f(H_0))\} \mid = 0.$$ 
Soit $$I_L= \frac{1}{\mud L^{d_1}}\ tr \{
\chi_{A_L\backslash A_{L,n}}(f(H)-f(H_0))\},$$
on choisit C >0 tel que $A_L\backslash A_{L,n}\subset B_L$, avec
$$B_L=\mathop\cap_{n=1}^{N_0}\{x\in A_L\setminus \mid \pi^nx \mid
\geqslant CL^{\alpha}\} ,$$   par cons{\'e}quent $$\mid I_L \mid \leqslant \frac{1}{\mud L^{d_1}} \mid tr \{ \chi_{B_L}(f(H)-f(H_0))\}\mid.$$
Or d'apr{\`e}s la Proposition \ref{pa1} on a  $$ \lim_{L\to \infty} \mid
I_L\mid =0.$$ 
D'o{\`u} $$  \lim_{ L \rightarrow \infty}\
\mid \nu_s^L(f)-\frac{1}{\mud L^{d_1}} \sum_{n=1}^{N_0}tr\{ \chi_{ A_{L,n}}
(f(H_n) - f(H_0))\} \mid=0.\qquad\di$$
\end{preuve}\par \bigskip
Pour chaque $n\in\{1,...,N_0\},$ on peut trouver un syst{\`e}me des
cordonn{\'e}es tel que $H_0= -\Delta +v_0$ s'{\'e}crit $\widetilde H_0 = \ds 
-\sum_{1\leqslant j\leqslant d }\frac{\partial^2}{\partial x_j^2
}+\widetilde v_0$ et $v_n(x)=\widetilde v_n(x',x'')$ avec $$\mid v_n(x',x'')\mid\leqslant
\frac{c_N}{1+\mid x'\mid^N},$$ pour tout $n\geqslant 1,$ 
$x'\in\R^{d-d_n}$ et $x''\in\R^{d_n}.$ Soit  $\widetilde H_n=
\widetilde H_0 +\widetilde v_n$ et $j_n: \ X_n\longrightarrow
\R^{d_n}$ un isomorphisme d'espaces  euclidiens sur
$\R^{d_n}$ et  d{\'e}finissons $$\widetilde \pi_n := j_n \circ\pi_n \quad
{\rm et} \quad
\widetilde \pi^n :=I_{\R^{d_n}} -j_n \circ\pi_n, $$ et $\widetilde A_{L,n}= \{\widetilde x \in \widetilde A_L,
  \mid \widetilde\pi^n x \mid \leqslant L^{\alpha }\}\ $ o{\`u}
  $\widetilde  A_L$  la boule habituelle de rayon L .
Pour $$(j,i)\in(\Z^{d-d_n}\times\Z^{d_n})\cap\widetilde A_L,$$ on  introduit la notation  suivante: $$\alpha_{j,i} = \alpha_{j,i}(f) = tr \{
\chi_{ B_{j,i}} (f(\widetilde H_n) - f(\widetilde H_o) ) \}$$ o{\`u}
$B_{j,i}$ est le cube unit{\'e} de centre
$(j,i)$. On aura besoin de la proposition
suivante qui sera d{\'e}montr{\'e}e dans le paragraphe \ref{deux} :
\begin{prop}\label{pd} Il existe une constante $d(f)>0$ telle que:
 $${\e} (\mid \alpha_{j,i} \mid ) \leqslant \frac {d(f)}{1 + \mid j\mid ^{2(d-d_n)}_1}, $$
o{\`u} $\mid j\mid_1 =j_1+j_2+...+j_{d-d_n}$ et $ {\e}(...)$ d{\'e}signe  l'esp{\'e}rance de $(\Omega ,\P) $.
\end{prop}\par \bigskip 
\pn {\bf Preuve du Th{\'e}or{\`e}me \ref{th}:}Le r{\'e}sultat du  Th{\'e}or{\`e}me $ 1.1$
d{\'e}coule de: \begin{equation} \label{aa} 
\widetilde\nu_s^{L,n}(f)=\frac{1}{\mudn L^{d_n}}tr\{ \chi_{ \widetilde
  A_{L,n}} (f(\widetilde H_n)
- f(\widetilde H_0)) \},
\end{equation} 
il suffit, alors de montrer (\ref{aa}) quand $L\to\infty .$ On peut remplacer $\chi_{\widetilde
  A_{L,n}}$ par $\chi_{\widetilde
  A_{L,n}'}$ o{\`u} $$\widetilde A_{L,n}'={\mathop\cup_{(j,i)\in\Z^d\cap\widetilde  A_{L,n}}}
  B_{(j,i)}$$ et $B_{(j,i)}$ est le cube  unit{\'e} de centre $(j,i)$
  . En effet $$\mu(\chi_{\widetilde
  A_{L,n}'}\setminus\chi_{\widetilde
  A_{L,n}})\leqslant cL^{d_n-1+\alpha(d-d_n)}$$ on d{\'e}duit que  
$$ \lim_{L\to\infty}\mid\widetilde\nu_s^{L,n}(f)-\frac{1}{\mudn L^{d_n}}
\sum_{(j,i) \in \Z^d \cap\widetilde  A_{L,n}} tr \{ \chi_{ B_{j,i}}
(f(\widetilde H_n) - f(\widetilde H_o))\}\mid=0.$$
Soit $\epsilon > 0 $, il existe M>0 tel que :$$ \sum_{\{(j,i) \in
  \Z^d \cap \widetilde A_{L,n} / \mid j \mid_1 \geqslant M \}}\frac {d(f)}{1 + \mid
  j\mid_1 ^{2(d-d_n)}} \leqslant \frac{\epsilon }{2}.$$  
On peut {\'e}crire alors, 
\begin{eqnarray}\label{1}
\widetilde\nu_s^{L,n}(f) & = & \frac{1}{\mudn L^{d_n}} \displaystyle  \sum_{(j,i) \in
  \widetilde A_{L,n} \cap {\Z^{d}}}   \alpha_{j,i} \nonumber\\ & = &
\frac{1}{\mudn L^{d_n}} \sum_{\{{(j,i)\in\widetilde  A_{L,n} \cap {\Z^{d}}} / \mid j
\mid_1 \leqslant M\} } \alpha_{j,i} + \frac{1}{\mudn L^{d_n}} \ \sum_{\{{(j,i)\in \widetilde A_{L,n} \cap {\Z^{d}}} / \mid j
\mid_1 \geqslant M\} }
 \alpha_{j,i} .
\end{eqnarray} 
En vertu de  la Proposition $\ref{pd}$, le deuxi{\`e}me terme de
l'expression  (\ref{1}) se majore par
$\epsilon/2$: 
\begin{eqnarray}
{\e}(\mid \frac{1}{\mudn L^{d_n}} \displaystyle \sum_{\{{(j,i)\in \widetilde A_{L,n} \cap {\Z^{d}}} / \mid j
\mid_1 \geqslant M\} }\alpha_{j,i} \mid )& \leqslant&\!\!\! \sum_{\{j \in
\Z^{d-d_n}/\mid j \mid_1 \geqslant M\}}
\!\!\!\!\!\e(
\mid \alpha_{j,i} \mid)\nonumber\\ & \leqslant& \!\!\!
\sum_{\{j \in
\Z^{d-d_n}/\mid j \mid_1 \geqslant M\}}\!\!
\frac{d}{1+\mid j\mid_1 ^{2(d-d_n)}}\leqslant \frac{\epsilon}{2}\nonumber,
\end{eqnarray} 
alors que le premier terme de l'expression (\ref{1}) converge
\begin{eqnarray} \label{e1}
 \frac{1}{\mudn L^{d_n}}\!\!\!\sum_{\{(j,i) \in \widetilde A_{L,n} \cap {\Z^{d}}/ \mid j
\mid_1 \leqslant M\}} \alpha_{j,i} & =  & \frac{1}{\mudn
L^{d_n}}\sum_{\mid j \mid_1 \leqslant M}\sum_{ \mid i\mid²+\mid j\mid² \leqslant
L²}\alpha_{j,i}\nonumber\\ &=&
\frac{1}{\mudn L^{d_n}}\sum_{\mid j\mid_1
  \leqslant M}  \sum_{i\in B(0,\sqrt{L-\mid j\mid²}) }
\alpha_{j,i}\nonumber\\ &\ds\mathop\sim_{L\to\infty}&\!\!\!\!\sum_{\mid j\mid_1
  \leqslant M}\!\!\!\frac{1}{\mudn (L-\mid j\mid²) ^{(d-d_n)/2}}\!\!\!\sum_{i\in B(0,\sqrt{L-\mid j\mid²}) }\!\!\!\!\!\!\!\alpha_{j,i}.\\
\nonumber
\end{eqnarray} 
Or en utilisant  le th{\'e}or{\`e}me ergodique de Birkhoff voir
$(\cite{Pastur},\cite{Carmona})$ on a l'existence de la limite de
(\ref{e1} ) si $L \to \infty$  presque partout et {\'e}gale {\`a} $\e(\alpha_{j,0})$.
Donc on a d{\'e}montr{\'e} l'existence de : $$\displaystyle\lim_{ L
  \rightarrow \infty}\widetilde \nu^{L,n}_s = \nu^{n}_s,$$on peut
conclure que $$ \displaystyle\lim_{ L
  \rightarrow \infty} \nu^{L,n}_s = \displaystyle\lim_{ L
  \rightarrow \infty} \frac{1}{\mudn L^{d_{n}}}tr\{ \chi_{ A_{L,n}} (f( H_n)
- f( H_0)) \} = \nu^{n}_s.$$
Par cons{\'e}quent  si $n\not\in {\cal K}=\{n\in\N,\ d_1=d_n\}$ on a : $$ \displaystyle 
\lim_{ L \rightarrow \infty}\frac{1}{\mud L^{d_{1}}}tr\{ \chi_{  A_{L,n}} (f( H_n)
- f( H_0)) \} = 0,$$ 
d'o{\`u} $$\displaystyle\lim_{L\rightarrow\infty}\nu_s^{L}=\displaystyle 
\lim_{L\rightarrow \infty}\frac{1}{\mud L^{d_{1}}}\sum_{n\in{\cal K}}tr\{\chi_{A_{L,n}}(f(H_1)
- f(H_0)) \}=\sum_{n\in{\cal K}}\nu_s^{n}.$$
Pour terminer la d{\'e}monstration du  Th{\'e}or{\`e}me $\ref{th},$ il reste {\`a} d{\'e}montrer la Proposition $\ref{pd}.\qquad\di$
\section{Preuve de la  Proposition \ref{pd}}\label{deux}
Dans cette partie, on va montrer un r{\'e}sultat
g{\'e}n{\'e}ral. On suppose que $v_1 $ est un potentiel dans ${\R^d},$ avec $ v_{1}$ ergodique par rapport {\`a}  $x''\in \R^{d-d_1}$  et  \
$$ 
\mid v_1(x',x'') \mid \leqslant \frac{c_N}{1+\mid x'\mid ^N}.$$ \\ On d{\'e}finit
$H_{1}=H_{0}+v_{1} $ avec $H_{0}$ le laplacien sur $\R^d$. Soient
$B_{j,i}\ \textsl{ pour}\  j \in {\Z^{d_1}},\ i\in {\Z^{d-d_1}} \ $  la boule     
$$B_{j,i}=\{(x',x'')\in{\R^d}/ \mid x'-j\mid²+\mid x''-i\mid² \leqslant c²\},$$
et$$\alpha_{j,i} = \alpha_{j,i}(f) = tr \{
\chi_{ B_{j,i}} (f(H_1) - f(H_o))\}.$$ Donc la Proposition $\ref{pd}$
r{\'e}sulte de la  proposition suivante: 
\begin{prop}\label{paa}
On suppose que $ f \in C^3 ({\R})$ et $ f^{(l)}(x) = 0(e^{- \alpha
  x})$ pour un certain \md\pn$( \alpha > 0 \ et\  l= 1,2,3 )\ {\rm si}\  x
\rightarrow\infty $. Alors il existe une constante $d (f)>0$ telle que
:
$${\e}(\mid \alpha_{j,i} \mid)  \leqslant \frac {d(f)}{1 + \mid
  j\mid^{2d_1}_1} $$\ o{\`u} $\mid j \mid_1 = j_1+j_2...+j_{d_1}$
et $\e$ d{\'e}signe l'esp{\'e}rance de $(\Omega, \P)$.
\end{prop}
On va commencer par le cas particulier $ f(x)= 0(e^{-tx})$ pour
$t>0.$ Puisque $H_1 \ {\rm et}\ H_0 $ sont born{\'e}s, on peut supposer 
que $H_1> 1\ {\rm et}\ H_0 > 1.$\\
Pour la preuve de la  Proposition $\ref{paa}$, on va utiliser les
deux  lemmes suivants: 
\begin{lem} \label{pbb}
Pour certains $ \beta , \gamma , \lambda > 0$,
\begin{equation*}\mid \alpha_{j,i}(e^{-tx}) \mid \quad\leqslant \lambda e^{- \gamma \mid j\mid_1^{d_1}}
e^{- \beta t}.
\end{equation*}
\end{lem}
\begin{lem}\label{pcc}
Soit $ \psi (z) = e^{- \beta z} \alpha_{j,i}(e^{-z}) $ une  fonction
complexe d{\'e}finie pour  $ (Rez \geqslant  0 ) $. On a alors :  $$\psi (t + is)
\leqslant \lambda e^{- \gamma \mid j\mid_1^{d_1}} {\rm arctan} \frac {t}{ \mid s \mid},$$
par cons{\'e}quent : $$ \mid \alpha_{j,i} (e^{-(t+is)}) \mid\leqslant
\lambda e^{-\beta t} e^{- \gamma \mid j \mid_1^{d_1}}{\rm arctan} \frac {t}{ \mid s \mid}.$$ 
\end{lem} 
La d{\'e}monstration de ces deux lemmes peut {\^e}tre trouv{\'e}e dans $(\cite{English})$.\md\pn{\bf Preuve de la Proposition \ref{paa}.}
Soit $f \in C^3 ({\R})$  et $ f(x) = 0(e^{- \alpha x })$ si $ x
\rightarrow \infty ( \alpha > 0) $ . \md \pn On peut {\'e}crire $$f(x) = \frac {1}{(x+c)^{2}}g(x)\qquad {\rm si}\quad x> c>0,$$
o{\`u} $g$ est de  m{\^e}me type que $f$. On a alors: $$g(x) = \int \widetilde{g}(s)
e^{-isx} ds.$$
Par cons{\'e}quent on a:
\begin{eqnarray}
 f(H_{1}) - f(H_{0})& = & (c+H_1)^{-2}g(H_{1}) -
(c+H_0)^{-2}g(H_{0}) \nonumber\\
& = & \int_{{\R}}((c+H_1)^{-2}e^{-isH_{1}} - (c+H_0)^{-2}e^{-isH_{0}})
    \widetilde{g}(s) \ ds  \nonumber\\
& = & \int_{{\R}} \widetilde{g}(s) \int_{0}^{\infty}
t(e^{-(t+is)(H_{1}+c)} -e^{-(t+is)(H_{0}+c)}) \ dt\ ds,\nonumber\\ 
\nonumber
\end{eqnarray}
il en r{\'e}sulte que 
\begin{eqnarray}\label{f}
\mid \alpha_{j,i}(f)\mid & = & \mid tr \{\chi_{ B_{j,i}} (f(H_1) -
f(H_o))\} \mid \nonumber\\
& = & \mid \int_{{\R}} \widetilde{g}(s) \int_{0}^{\infty}
t\  tr\{ \chi_{c_{j,i}}(e^{-(t+is)(H_{1}+c)} -e^{-(t+is)(H_{0}+c)})\} \ dt\ ds\mid\nonumber\\
& \leqslant & \int_{{\R}} \widetilde{g}(s) \int_{0}^{\infty} t \ \mid
\alpha_{j,i}(e^{-(t+is)(x+c)}) \mid \ dt\  ds\nonumber\\
& \leqslant & \int_{{\R}} \widetilde{g}(s) \int_{0}^{\infty} t e^{-
  \beta t} e^{- \gamma \mid j\mid_1^{d_1} {\rm arctan} \frac {t}{\mid
  s \mid}}dt\quad ds.
\end{eqnarray}
Or$$ \int_{0}^{\infty} t e^{-
  \beta t} e^{- \gamma \mid j\mid_1^{d_1} {\rm arctan} \frac {t}{\mid s \mid}} \ dt \leqslant\frac {K(1 + s^2)}{1+\mid j\mid_1^{2d_1}}.$$\\
On conclut que le membre droit (\ref{f}) peut {\^e}tre estim{\'e} par  
$$  \frac{D_1}{1+\mid j\mid_1^{2d_1}}\int \mid  \widetilde{g}(s) \mid
(1+s^2) \ ds \leqslant \frac{d}{1+\mid j\mid_1^{2d_1}},$$\\ 
 o{\`u} $d$ d{\'e}pend seulement de la fonction $g$ et par cons{\'e}quent de la fonction $f.\qquad\di$

\section{Appendice}\label{quatre}
Pour simplifier les preuves, on prend $t\in]0,1]$ le long  de ce  paragraphe.
\begin{prop}\label{paaa}
Pour tout $\chi \in C_0^{\infty}({{\R}^d}),$ et $\mid f(x) \mid
\leqslant c
e^{-x};$   
$$ \mid \mid \chi f(H) \mid \mid_{B_{1}} \leqslant c(f) \mu(\chi),$$ o{\`u}
$\mu(\chi) =\#\{ j\in\Z^d:\ {\rm supp}\chi\cap {\cal C}_j\neq\emptyset \}
$ et ${\cal C}_j=j+[\frac{-1}{2}\frac{-1}{2}]^d$.
\end{prop}
Pour montrer la Proposition \ref{paaa} nous utilisons les lemmes suivants:

\begin{lem}\label{pbbb}
Pour tout $N\in\N$ il existe une constante $C_{N}>0$ telle que pour tout $n, k\in{\Z^d}$ on ait: $$\mid\mid
\chi_{n}e^{-tH}\chi_{k}\mid \mid_{B_2}\leqslant
\frac{C_N}{t^{d/2}(1+\mid n-k\mid^N)}.$$
\end{lem}
\begin{preuve}
D'apr{\`e}s l'article de Davies  $(\cite{Davies}) $ on a : $$e^{-tH}\varphi(x)=\int K_t
(x,y)\varphi(y) dy, $$ avec $ 0\leqslant K_t(x,y) \leqslant ct^{-d/2}e^{-a\mid
  x-y \mid^2}.$ Par suite 
\begin{equation}\label{midb}
 \mid\mid
\chi_{n}e^{-tH}\chi_{k} \mid
\mid_{B_2}^2=\int\!\!\!\int\mid \chi_{n}(x)\chi_{k}(y) K_t
  (x,y)\mid^2dxdy \leqslant Ct^{-d}e^{-\alpha\mid n-k\mid^2} ,
\end{equation}
or pour tout $N\in\N$ il existe une constante $C_{N}>0$ telle que $e^{-\alpha\mid n-k\mid^2}\leqslant\frac{C_N}{1+\mid n-k\mid^N}
$, alors on peut majorer (\ref{midb}) comme enonc{\'e} dans le Lemme
\ref{pbbb}. $\quad \di $
\end{preuve}\md\pn

\begin{lem}\label{pbbb'}
Pour tout $N\in\N$ il existe une constante $C_{N}>0$ telle que pour tout $n, k\in{\Z^d}$ on ait: $$\mid\mid
\chi_{n}e^{-tH}\chi_{k}\mid \mid_{B_1}\leqslant
\frac{C_N}{t^{d}(1+\mid n-k\mid^N)}.$$
\end{lem}
\begin{preuve}\label{pb'} 
En utilisant la partition de l'identit{\'e} on  {\'e}crit
\begin{eqnarray}
\mid\mid
\chi_{n}e^{-tH}\chi_{k}\mid \mid_{B_1} & =&\sum_{m\in\Z^d}\mid\mid
\chi_{n}e^{-tH/2}\chi_m^2e^{-tH/2}\chi_{k}\mid \mid_{B_1}\nonumber\\
&\leqslant&\sum_{m\in\Z^d}\mid\mid\chi_{n}e^{-tH/2}\chi_m\mid\mid_{B_2}\mid\mid\chi_m
e^{-tH/2}\chi_{k}\mid \mid_{B_2}\nonumber\\ &\leqslant&\sum_{m\in\Z^d}\frac{c_N}{t^{d/2}(1+\mid n-m\mid^N)}\frac{c_N}{t^{d/2}(1+\mid k-m\mid^N)}\nonumber\\&\leqslant&\frac{C_N}{t^{d}(1+\mid k-n\mid^N)}.\quad \di \nonumber
\end{eqnarray}
\end{preuve}
\begin{lem}\label{pddd}Soit $n \in {\Z}$ et soit $ w$ une fonction mesurable
  telle que $ 0\leqslant w \leqslant 1\
  $. On pose $L=dist(suppw,n)+1$. Alors pour tout $N\in \N$ il existe
  $C_N$ ind{\'e}pendant de $w$ et $n$ tel que :$$\mid \mid \chi_{n}e^{-tH}w \mid
\mid_{B_1} \leqslant \frac{C_N}{t^{d} L^N}.$$
\end{lem}
\begin{preuve} 
En utilisant le fait que $$w=w\sum_{\{m\in\Z^d/\mid m-n\mid\geqslant
  L\}}\chi_m,$$ on a l'estimation suivante: 
\begin{eqnarray}
\mid \mid \chi_{n}e^{-tH}w \mid
\mid_{B_1} &\leqslant& \sum_{\{m\in\Z^d/\mid m-n\mid\geqslant L\}}\mid \mid \chi_{n}e^{-tH}\chi_m\mid
\mid_{B_1} 
  \leqslant\sum_{\{m\in\Z^d/\mid m-n\mid\geqslant
    L\}}\frac{c_{2N}}{t^{d}(1+\mid  n-m\mid^{2N})}\nonumber\\
  &=&\sum_{\{k\in\Z^d/\mid k\mid\geqslant
    L\}}\frac{c_{2N}}{t^{d}(1+\mid
    k\mid^{2N})}\leqslant\sum_{\{k\in\Z^d/\mid k\mid\geqslant
    1\}}\frac{c_{2N}}{t^{d}(1+\mid k\mid^{N}L^{N})}\nonumber\\ &\leqslant&\frac{c_{2N}}{L^{N}t^{d}}\sum_{\{k\in\Z^d/\mid k\mid\geqslant 1\}}\frac{1}{\mid k\mid^{N}}\leqslant\frac{C_{2N}}{t^{d} L^N}.\nonumber
\end{eqnarray}
\end{preuve}
Ainsi la d{\'e}monstation est achev{\'e}e .$\qquad \di$\par \bigskip
\pn{\bf Preuve de la Proposition \ref{paaa}.}: En utilisant la partition de l'identit{\'e} 
$$I= \sum_{n \in{\Z^d}} \chi_{n}^2,\quad{\rm et}\quad\chi=\chi\sum_{j\in
  J}\chi_j$$ o{\`u} $J = \{ j\in\Z^d:\ {\rm supp}\chi\cap{\cal  C}_j\neq\emptyset \},$
on {\'e}crit 
$$ \chi e^{-tH} =\chi\sum_{n \in{\Z^d},j\in J}\chi_j e^{-H}\chi_n.$$
Donc il existe $c_0$ tel que pour tout $n\in\Z^d, j\in J$ on a\\
1+dist(supp$\chi_j$, supp$\chi_n)\geqslant c_0(1+\mid n-j\mid)$, par
cons{\'e}quent 
\begin{eqnarray}
\mid \mid \chi e^{-tH} \mid \mid_{B_{1}} &\leqslant& \sum_{n
  \in{\Z^d},j\in J}\mid \mid \chi_j e^{-tH}\chi_{n}\mid \mid_{B_1} \nonumber\\
&\leqslant& \sum_{n \in{\Z^d}, j\in J} \frac{c_N c }{ (1+\mid
  n-j\mid)^Nt^{d}}\nonumber\\ &\leqslant&\sum_{j\in J}
\frac{C_Nc}{t^{d}}\leqslant C_Nct^{-d}\mu(\chi).\nonumber
\end{eqnarray}
Si $\mid f(x)\mid\leqslant c e^{-x} $ il existe $g \in
L^{\infty}({\R})$\ telle  que $ f(x)=e^{-x}g(x).$ 
On conclu :$$\mid \mid \chi f(H)\mid \mid_{B_1}=\mid \mid \chi e^{-H}g(H)\mid
\mid_{B_1} \leqslant\mid \mid g(H) \mid \mid\  \mid \mid \chi e^{-H}
\mid \mid_{B_1}\leqslant c(f) \mu(\chi).\qquad\di$$
\begin{lem}\label{peee}Soit $k\in\{1,...,N_0\}$ et $w_L$ définit comme
  dans la preuve du Lemme \ref{pc} avec n=1. Alors 
pour tout $N\in\N$ il existe une constante $C_N>0$ telle que
$$\mid\mid v_k\chi_{\{x\in A_{4L})
  / \mid\pi^kx\mid>L^{\alpha}\}}w_Le^{-tH}\mid\mid_{B_1}\leqslant C_N t^{-d}L^{-N}.$$
\end{lem}
\begin{preuve}On a
\begin{eqnarray}\label{4} 
\mid\mid v_k\chi_{\{x\in A_{4L}
  /  \mid\pi^kx\mid>L^{\alpha}\}}w_Le^{-tH}\mid\mid_{B_1}\!\!\!\!\!&\leqslant&\!\!\!\!\!\mid\mid v_k\chi_{\{x\in A_{4L}  /\mid\pi^kx\mid>L^{\alpha}\}}\mid\mid\ \mid\mid w_{4L}e^{-tH}\mid\mid_{B_1}\nonumber\\
&\leqslant&\!\!\!C_{N'}(1+\mid
\pi^kx\mid)^{-N'}\mid\mid \chi_{A_{L}}e^{-tH}\mid\mid_{B_1}\nonumber\\&\leqslant&C_{N'}(1+
L^{\alpha})^{-N'}cL^{d}t^{-d}\leqslant C_N t^{-d}L^{-N} \nonumber. \qquad\di
\end{eqnarray}
 
\end{preuve}
\begin{lem}\label{peee'} Soit $w_L$ comme dans la preuve du Lemme
  \ref{pc} avec n=1,
pour tout $k\in\{2,...,N_0\},$ il existe une constante $c>0$ telle que
$$\mid\mid v_k\chi_{\{x\in A_{4L}    / \mid\pi^kx\mid<L^{\alpha}\}}w_Le^{-tH}\mid\mid_{B_1}\leqslant
c t^{-d}L^{d_1-1+\alpha d}.$$
\end{lem}
\begin{preuve}
Tout $x\in X $  peut s'{\'e}crire de fa{\c c}on unique de la
forme  $x = x'+x''$ o{\`u} $x'\in X_{1} \cap X_{k} $ et $  x''\in (X_{1}\cap
 X_{k})^{\bot}.$  Or  si de plus $x\in A_{4L}$ donc $\mid x'\mid\leqslant
 4L$ et en utilisant le fait que   dist$(x, X_{k})\leqslant L^{\alpha}$ et
dist$(x,X_{1})\leqslant
 cL^{\alpha}$ si $x\in$ supp$w_L$ donc $\mid x''\mid= {\rm
  dist}(x,X_{1}\cap X_{k}) \leqslant c L^{\alpha}$. Par cons{\'e}quent 
\begin{eqnarray}\label{4'}
\mid\mid v_k\chi_{\{x\in A_{4L}
  / \mid\pi^kx\mid<L^{\alpha}\}}w_Le^{-tH}\mid\mid_{B_1}\!\!&\leqslant&\!\!\!\!
 c\mid\mid\chi_{\{x\in A_{4L}
  / \mid x''\mid<cL^{\alpha}\}}e^{-tH}\!\!\mid\mid_{B_1}.
\end{eqnarray}
Si on pose $d_{k,1}=dim (X_{1}\cap X_{k}) $ on remarque que
$d_{k,1}\leqslant d_1-1$. Ainsi  en utilisant la Proposition \ref{paaa} et le fait que $\mu(\chi_{\{x\in A_{4L}
  / \mid x''\mid<cL^{\alpha}\}})\leqslant
cL^{d_{k,1}+\alpha(d-d_{k,1})},$ on a le membre de gauche de (\ref{4'})
est major{\'e} par $$ C t^{-d}L^{d_{k,1}+\alpha(d-d_{k,1})}\leqslant C
t^{-d}L^{d_{1}-1+\alpha d}.\qquad\di$$
\end{preuve}
\begin{cor}\label{pfff}
On a l'estimation $(\ref{c'}) $ et $(\ref{b'})$.
\end{cor}
\begin{preuve}En utilisant la partition de l'identit{\'e} $$ I=
  \chi_{\{x\in A_{4L} 
  / \mid\pi^kx\mid<L^{\alpha}\}}+
\chi_{\{x\in A_{4L}
  / \mid\pi^kx\mid>L^{\alpha}\}},$$ en vertu du Lemme \ref{peee} et
\ref{peee'} on a  
\begin{eqnarray}
\mid\mid v_k w_L
e^{-tH}\mid\mid_{B_1}&\leqslant& \mid\mid v_k\chi_{\{x\in A_{4L} 
  / \mid\pi^kx\mid<L^{\alpha}\}} w_L
e^{-tH}\mid\mid_{B_1}\nonumber\\&&+\mid\mid v_k\chi_{\{x\in A_{4L} 
  / \mid\pi^kx\mid>L^{\alpha}\}} w_L
e^{-tH}\mid\mid_{B_1}\nonumber\\ &\leqslant&c_Nt^{-d}L^{-N}+ C
t^{-d}L^{d_{1}+\alpha d}.\nonumber
\end{eqnarray}
Par cons{\'e}quent  on a $$\mid\mid\sum_{k=2}^{N_0} v_k w_L
e^{-tH}\mid\mid_{B_1}\leqslant\sum_{k=2}^{N_0}\frac{c(L^{\alpha d+d_1-1}+L^{-N})}{t^{d}}
\leqslant C\frac{(L^{\alpha d+d_1-1}+L^{-N})}{t^{d}},$$
d'o{\`u} l'estimation  $(\ref{c'}) .$ 
De la m{\^e}me fa{\c c}on, on montre que   (\ref{b'}) a lieu. \qquad $\di$ \par \bigskip
\end{preuve}
Dans cette partie on va montrer le lemme \ref{pc'}, pour cela on a
besoin de ces deux 
 lemmes suivants.
\begin {lem}\label{phhh}Pour tout $\epsilon >0 , N\in\N,$ il existe une constante  $C_{N}>0$
telle que:$$ \mid\mid \chi_k
e^{-tH}(1-\chi_{B(k,L^{\epsilon})})\mid\mid_{B_1}\leqslant
C_{N} L^{-\epsilon N}
t^{-d}.$$
\end{lem}
\begin{preuve}
En utilisant  $$1-\chi_{B(k,L^{\epsilon})} = \sum_{\{ m\in \Z^d, dist(k,m)>L^{\epsilon}\}} \chi_m(1-\chi_{B(k,L^{\epsilon})}),$$
on a  
\begin{eqnarray}
\mid\mid \chi_k
e^{-tH}(1-\chi_{B(k,L^{\epsilon})})\mid\mid_{B_1}&\leqslant& \sum_{\{
  m\in \Z^d,  dist(k,m)>L^{\epsilon}\}} \mid\mid\chi_k e^{-tH}\chi_m\mid\mid_{B_1}\nonumber\\
&\leqslant & \sum_{\{m\in \Z^d, dist(k,m)>L^{\epsilon}\}} \frac{c_N
  t^{-d}}{1+\mid k-m\mid^N}\leqslant\sum_{\mid m\mid\geqslant L^{\epsilon}}\frac{c_N
  t^{-d}}{1+\mid m\mid^N}\nonumber\\ &\leqslant&\sum_{\mid m\mid\geqslant L^{\epsilon}}\frac{c_N
  t^{-d}}{1+(L^{\epsilon/2}\sqrt{ m})^N}\leqslant\sum_{\mid
  m\mid\leqslant 1}\frac{c_N
  t^{-d}}{(L^{\epsilon/2}\sqrt{m})^N}
\nonumber\\ &\leqslant&C_{N}L^{-\epsilon N/2}t^{-d} .\qquad\di\nonumber
\end{eqnarray}
\end{preuve}
\begin{lem}\label{pjjj}
Soit $L_{k} =  dist(k,\ds\cup_{n=1}^{N_0} X_n)+1.$ Alors pour tout $N\in\N ,
$ il existe une constante $C_N>0$ telle que 
\begin{equation} \label{1a}
\mid\mid \chi_k
(e^{-tH} - e^{-tH_n})\mid\mid_{B_1} \leqslant C_N L_k ^{-N/2} t^{-d}
\end{equation}
\end{lem}
\begin{preuve}Pour simplifier la d{\'e}monstration on prend $n=0$.
On va utiliser la r{\'e}currence par rapport {\`a} $N\in\N$. Pour $N = 0$, on a toujours $\mid\mid \chi_k(e^{-tH} -
e^{-tH_0})\mid\mid_{B_1} \leqslant c_o t^{-d}$. On {\'e}crit $\chi_k(e^{-tH} - e^{-tH_0}) =
J_1+J_2$ avec $$J_1= \chi_k(e^{-tH}-e^{-tH/2 }e^{-tH_0/2})$$
et $$J_2= 
\chi_k(e^{-tH/2 }e^{-tH_0/2}-e^{-tH_0}).$$
Montrons  tout d'abord que pour tout $N\in\N$ on a
\begin{equation}\label{1c}
\mid\mid J_1\mid\mid_{B_1}\leqslant C'_N L_k^{-N/2}  t^{-d}.
\end{equation} 
Par r{\'e}currence  on suppose (\ref{1c}) est vraie pour
$N$ et on  montre qu'il est  vraie  pour $N+1$. Soit $\widetilde \chi_k$ une fonction lisse telle que
$\mid\partial^{\alpha}\widetilde\chi_k\mid\leqslant c_{\alpha}L_k^{-\mid \alpha\mid}$
pour tout $\alpha\in \N^d$ avec $$\widetilde
\chi_k =1\quad{\rm si}  \quad \mid x-k\mid
\leqslant\frac{L_k}{2}\quad{\rm et}\quad \widetilde\chi_k =0\quad{\rm si}  \quad \mid x-k\mid
\geqslant L_k.$$
On peut ecrire $$J_1= \chi_k( e^{-tH}-
e^{-tH_0})\widetilde\chi_k-\chi_k(e^{-t H/2}-e^{-t
  H_0/2})\widetilde\chi_k  e^{-tH_0/2}+R_k(t),$$
où en vertu du Lemme \ref{phhh} on a $\mid\mid R_k(t)\mid\mid_{B_1}\leqslant C_M L_k^{-M}t^{-d}$ pour
tout $M\in \N.$  
Par cons{\'e}quent  si on note  $ \widetilde{V}=\ds\sum_{n=1}^{N_0}v_n$ on a 
\begin{eqnarray}
 J_1&=& \int_{\frac{t}{2}}^t
 \frac{d}{ds}[\chi_k(e^{-sH}-e^{-sH_0})\widetilde\chi_ke^{-(t-s)H_0}]ds +R_k(t)\nonumber\\ &=&\int_{\frac{t}{2}}^t\chi_k(e^{-sH_0}-e^{-sH})[\Delta, \widetilde\chi_k]e^{-(t-s)H_0}+\chi_ke^{-sH}\widetilde {V}\widetilde\chi_{k}e^{-(t-s)H_0}ds+R_k(t).\nonumber
\end{eqnarray}
d'o{\`u}
\begin{eqnarray}\label{1b}
\mid\mid J_1\mid\mid_{B_1} & \leqslant& \!\!\!\int_{\frac{t}{2}}^t\!\!\!\mid\mid\chi_k(e^{-sH_0}-e^{-sH})\mid\mid_{B_1}\mid\mid[\Delta,\widetilde\chi_k]e^{-(t-s)H_0}\mid\mid
ds\nonumber\\ & &+\int_{\frac{t}{2}}^t\mid\mid\chi_k e^{-sH}\mid\mid_{B_1}\mid\mid\widetilde {V}\widetilde\chi_k
e^{-(t-s)H_0}\mid\mid ds +\mid\mid R_k(t)\mid\mid_{B_1}.
\end{eqnarray}
En utilisant l'hypoth{\`e}se de r{\'e}currence et   $\mid \mid
[\Delta,\widetilde\chi_k]e^{-(t-s)H_0}\mid\mid\leqslant
cL_k^{-1}(t-s)^{-1/2}$ 
on a le premier terme de (\ref{1b}) est major{\'e} par 
\begin{eqnarray}
& &\int_{\frac{t}{2}}^t C_N s^{-d}L_k^{-N/2}
cL_k^{-1}(t-s)^{-1/2} ds \leqslant C_N
ct^{-d}L_k^{-N/2-1}\int_{\frac{t}{2}}^t (t-s)^{-1/2} ds\nonumber\\ & &\leqslant C_{N+1}t^{-d}L_k^{-(N+1)/2}.\nonumber
\end{eqnarray}
Comme $\mid \widetilde{V}(x)\widetilde\chi_k(x) \mid \leqslant
c_{\gamma} L_k^{-\gamma}$ pour tout $\gamma$ et en vertue du Lemme
\ref{pb} on peut majoré le deuxième terme de manière semblable  et 
 conclure  que 
$$
\mid\mid J_1\mid\mid_{B_1} \leqslant 3C_{N+1}L_k^{-(N+1)/2} t^{-d}.$$ 
De  fa{\c c}on similaire, en utilise la récurrence pour montrer que
pour tout $N\in\N$ on a  
$$\mid\mid J_2\mid\mid_{B_1}\leqslant C'_N L_k^{-N/2} t^{-d}.$$ 
En effet, en {\'e}crivant  
\begin{eqnarray}
J_2 &= &\chi_k\widetilde\chi_k( e^{-tH_0}-
e^{-tH})\chi_{k,\epsilon}-\chi_{k}e^{-tH/2}\widetilde\chi_k(e^{-tH_0/2}-e^{-tH/2})\chi_{k,\epsilon}
+\widetilde R_k(t)\nonumber\\ &=&\int_{\frac{t}{2}}^t
 \chi_k\frac{d}{ds}[e^{-(t-s)H}\widetilde\chi_k  (e^{-sH_0}
 -e^{-sH})\chi_{k,\epsilon}]ds +\widetilde R_k(t),\nonumber
\end{eqnarray}
o{\`u} $\chi_{k,\epsilon}$ est une fonction {\`a} support dans
$B(k,L_k^{\epsilon})$ et par le Lemme \ref{phhh} on a\md\pn
$\mid\mid\widetilde R_k(t)\mid\mid_{B_1}\leqslant C_M L_k^{-M}t^{-d}$.
Par cons{\'e}quent
\begin{eqnarray}
\mid\mid J_2 \mid\mid_{B_1}&\leqslant&
\int_{\frac{t}{2}}^t\!\!\!\mid\mid(e^{-sH_0}-e^{-sH})\chi_{k,\epsilon}\mid\mid_{B_1}\mid\mid
 e^{-(t-s)H}[\Delta,\widetilde\chi_k]\mid\mid
ds\nonumber\\ & &+\int_{\frac{t}{2}}^t\mid\mid\chi_k
e^{-(t-s)H}\mid\mid_{B_1}\mid\mid \widetilde V\widetilde\chi_k e^{-sH}\mid\mid ds +\mid\mid
\widetilde R_k(t)\mid\mid_{B_1}\nonumber\\ &\leqslant& \int_{\frac{t}{2}}^t 2C_N
L_k^{-N/2-1+\epsilon d}(t-s)^{-1/2}s^{-d} ds + C_N
L_k^{-N/2-1}t^{-d}\nonumber\\&\leqslant& 3C_{N+1}L_k^{-N/2-1+\epsilon
  d}t^{-d}\leqslant 3C_{N+1}L_k^{-N/2-1/2}t^{-d}\nonumber \quad{\rm si}\quad
\epsilon d\leqslant \frac{1}{2}.
\end{eqnarray}
Donc $$\mid\mid \chi_k(e^{-tH}-e^{-tH_0})\mid\mid_{B_1}\leqslant
C'_{N+1} L_k^{-(N+1)/2}t^{-d}.$$ Ainsi l'assertion du lemme est vraie pour $N+1$
et la d{\'e}monstration est achev{\'e}e. $\qquad\di$ \par \bigskip \pn
{\bf Preuve du Lemme \ref{pc'}.:}

 Comme
$$\chi_{A_{L,n}}(1-\chi_{\epsilon})=\chi_{A_{L,n}}(1-\chi_{\epsilon})\sum_{k\in K} \chi_k$$ o{\`u} $K=
\{{k\in\Z^d,dist(k,\cup_n X_{n=1}^{N_0})\geqslant
  L^{\epsilon}et\mid k\mid<L}\},$
on peut conclure par le Lemme \ref{pjjj} que 
$$\mid\mid\chi_{A_{L,n}}(1-\chi_{\epsilon})(e^{-tH}-e^{-tH_n})\mid\mid_{B_1}\leqslant
\sum_{k\in K}\mid\mid\chi_k
(e^{-tH}-e^{-tH_n})\mid\mid_{B_1}\leqslant C_N
L^{-\epsilon N/2+d} t^{-d}$$ et pour $N$ suffisament grand, on obtient $$ \lim_{ L \rightarrow
  \infty}\ \frac{1}{\mud L^{d_1}}  
\mid\mid\chi_{A_{L,n}}(1-\chi_{\epsilon})(e^{-tH}-e^{-tH_n})\mid\mid_{B_1}=0.\qquad\di$$

\end{preuve}
{\bf Remerciements:}\\Je remercie tout d'abord M Lech Zielinski pour
son aide et  ses critiques pendant toute la dur{\'e}e de mon travail. Je
remercie aussi A{\"\i}cha pour sa collaboration dans le domaine de l'informatique et Aladin pour
sa patience . 
\goodbreak

\pn 
{\footnotesize{\begin{center}
{Boutheina {\sc Souabni}:\md\pn Laboratoire de math{\'e}matique \md\pn et physique.
Institut de \md\pn Math{\'e}matiques de Jussieu, \md\pn F-75251 Paris, France.}\md\pn\email{ Boutheina@math.jussieu.fr }
\end{center}}}
\end{document}